\documentclass[prl,twocolumn]{revtex4}
\usepackage[english]{babel}
\usepackage{bm}
\usepackage{amssymb}
\usepackage{mathtools}
\usepackage{amsmath}
\usepackage{relsize}
\usepackage{hyperref}
\usepackage[mathscr]{euscript}
\usepackage{enumitem}
\usepackage{csquotes}
\usepackage{natbib}
\begin{document}

\title{Geometrization of the scalar matter}
\author{Avadhut V. Purohit}
\email{avdhoot.purohit@gmail.com\\}
\affiliation{Chennai Mathematical Institute, H1 SIPCOT IT Park, Kelambakkam, Chennai, India - 603103}
\date{\today}

\begin{abstract}
	The scalar matter and gravity are unified into the geometric scalar matter and quantized. The quantum with a definite 3-metric has definite energy but does not have well-defined momentum. The quantum theory resolves singularities.
\end{abstract}

\maketitle
\section{Introduction}
There are primarily two major approaches to quantum gravity. The first approach looks for the unified field theory that includes gravity, and the second approach looks for the space theory of matter \citep{clifford}. This paper takes the second approach, which is also known as `\textit{geometrization of matter}' (refer to the corresponding section in \citep{odenwald} for a detailed discussion). 
\begin{displayquote}
    ``The material particle has no place as a fundamental concept in field theory. Even Maxwell's electrodynamics is not complete for this reason. Gravity as a field theory must also deny a preferred status to matter.''\\ 
    \hspace*{4.3cm}\rm --- Einstein, 1950
\end{displayquote}
The moment we identify a particular particle with a certain geometry, the occurrence of a virtual multi-particle state means the presence of a virtual multi-geometric state in the quantum evolution leading to the concept of `geometric field.' To achieve that, we need ADM theory \citep{adm} as it brings gravity closer to matter fields by casting it as a gauge theory. The ADM Hamiltonian constraints for \textit{pure} gravity then re-interpreted as a classical field equation.\\
\hspace*{3mm}Although \citep{ThirdQuantization}, \citep{pinbox}, \citep{rubakov}, \citep{strominger} and several others attempted such re-interpretation, all these approaches include the matter field along with the gravitational field. Conceptually, if already matter field $\phi$ exists in $\Phi(\phi, q_{ab})$ then what is $\Phi$? In such theories, $\Phi$ usually describes a particular Universe. The scope of such theories is Multiverse. Mathematically, such quantization is the second quantization for the gravitational field. But it is the third quantization for matter fields.\\
\hspace*{5mm}Mathematically, the geometric scalar field theory at the classical level is similar to the theory presented in \cite{ThirdQuantization} (with no matter field). But matter field is not included in the ADM constraints because matter itself is a geometric entity. Quantum mechanically, it is the second quantization and not the third quantization. The vacuum is a sea of constantly creating, annihilating quanta of the geometric scalar matter. The geometric scalar field theory resolves singularities.\\
\hspace*{5mm}The third section discusses the classical theory. The fourth section analyzes two applications, namely FLRW models of the Universe, Schwarzschild geometry, and linearized theory. The fifth section is devoted to quantum theory, followed by two applications, the flat Universe and Schwarzschild space-time. The last section discusses the resolution of issues in quantum gravity. The section also compares the theory with Wheeler-DeWitt theory, third quantization theories, pure gravity, and standard matter.\\
\hspace*{5mm}Note: \textit{$R^{(3)}$ is the Ricci scalar on 3-manifold of the ADM theory. $(\zeta, \zeta^{A})$ is the DeWitt superspace.}
\section{Motivation}
The general theory of relativity is a dynamic theory of space-time. Although the space-time is smooth, the matter fields are quantum at the micro-scale. Therefore, gravity needs to get quantized. There are two types of quantization. Namely, particle quantization and field quantization. The non-relativistic particle is quantized by taking $\vec{p} \rightarrow -i\vec{\nabla}$ and $E\rightarrow i\frac{\partial}{\partial t}$. But if we attempt to quantize the relativistic particle, implementing a similar procedure raises severe issues. The concept of `field' overcomes these difficulties. This is done by re-interpreting $E^2 = p^2 + m_0^2 \rightarrow \left(\partial_\mu\partial^\mu + m_0^2\right) \phi = 0$. The field $\phi$ is still classical even though we have used $\vec{p} \rightarrow -i\vec{\nabla}$ and $E\rightarrow i\frac{\partial}{\partial t}$. In the general theory of relativity, a certain matter configuration has a particular geometry $g_{\mu\nu}(\phi)$. But the quantum evolution has intermediate virtual multi-geometric states for a particular matter field. It forces us to the concept of a `geometric field.'\\
\hspace*{3mm}Having developed on the platform of special relativity, the standard fields such as Klein-Gordon field carry the same concept of particle. Whereas in the general theory of relativity, it is well-known from certain gravitational effects such as the Unruh effect that the standard concept of particle and vacuum of standard fields are not well-defined. Hence geometric matter fields cannot be expected to have mass. Particles have shape, energy, momentum, and mass. The geometric matter fields also have energy, momentum, and intrinsic geometry described by $\mu^2$ (see (\ref{coupling})) plays a role similar to the mass. In this way, matter and gravity get unified while implementing `geometric quantization.'
\section{Classical Theory}
\hspace*{3mm}I define the action functional for the geometric scalar field that gives field equation(\ref{FieldEqn}).
\begin{eqnarray}\label{ScalarAction}
  \mathcal{A}_{\Phi} \coloneqq \int \, \mathscr{D}\zeta \frac{\sqrt{-G}}{2} \left( \partial_{\mu}\Phi G^{\mu\nu}\partial_{\nu}\Phi - \mu^2 \Phi^{2} \right)
\end{eqnarray}
$\mathscr{D}\zeta$ is suitable measure over 6D manifold. $\sqrt{-G}\coloneqq \sqrt{-\text{det }G_{\mu\nu}}$ with 
\begin{eqnarray} 
& G_{\mu\nu} \coloneqq
  \begin{pmatrix}
    1  &0 \\
    0  &-\frac{3 \zeta^{2}}{32}\bar{G}_{AB}
  \end{pmatrix} \\ 
  &\bar{G}_{AB} \coloneqq \text{Tr}\left(\mathbf{q}^{-1}\frac{\partial \mathbf{q}} {\partial\zeta^A}\mathbf{q}^{-1}\frac{\partial \mathbf{q}}{\partial\zeta^B}\right) \\ \nonumber
    &\mathbf{q} \coloneqq q_{ab}
\end{eqnarray}
$\bar{G}_{AB}$ is symmetric DeWitt metric on 5D manifold $M$ identified with $SL(3, \mathbb{R})/SO(3, \mathbb{R})$ (refer \citep{DeWitt} and \citep{Giulini} for more discussion on the geometry of superspace).
\begin{equation}\label{coupling}
    \mu^2 \left(\zeta,\zeta^{A}\right)\coloneqq - \frac{3\zeta^{2}}{32} R^{(3)}
\end{equation}
act as a coupling parameter for the geometric scalar field.  
\begin{align}
    & \partial_{\mu} \coloneqq \left( \frac{\partial }{\partial \zeta}, \frac{\partial}{\partial \zeta^{A}} \right) & \, \zeta^{\mu} \coloneqq \left( \zeta, \zeta^{A} \right)
\end{align}
form a six-dimensional vector space with $A = 1, 2, 3, 4, 5$. Components of the 3-metric are `good' coordinates. $\zeta^A$ are chosen as components of the 3-metric. These are explicitly defined and discussed in great detail in the DeWitt's paper \citep{DeWitt}. Since $P^{ab}\rightarrow -i \frac{\partial}{\partial q_{ab}}$,
\begin{equation}
    D_{a}P^{ab} \approx 0 \Rightarrow D_{a}\frac{\partial \Phi}{\partial q_{ab}} \approx 0
\end{equation}
$\Phi$ and therefore $\frac{\partial \Phi}{\partial q_{ab}}$ being function of the 3-metric only, it naturally satisfies diffeomorphism constraints. Invariance of the action under variation of $\Phi$ gives the field equation as
\begin{eqnarray}\label{FieldEqn}
  \left(\frac{1}{\sqrt{-G}}\partial_\mu \sqrt{-G}G^{\mu\nu}\partial_\nu + \mu^2\right) \Phi = 0
\end{eqnarray}
This equation defines the field. The ADM theory does not have an issue with operator ordering. But for the classical geometric scalar field, different combinations of $P^{ab}$ and $q_{ab}$ will give non-equivalent field equations. I have taken the combination of field variables with a consistent self-adjoint extension. In other words, the combination allows the Hamiltonian operator to be self-adjoint (the quantum theory not done in this paper). On single spacetime-like interpretation $\Phi = e^{i q_{\mu\nu}P^{\mu\nu}}$, ADM Hamiltonian constraints for pure gravity in the DeWitt coordinates (5.20, \cite{DeWitt})
\begin{equation}
  P_0^2 - \frac{32}{3 \zeta^2}\bar{G}^{AB}P_A P_B + \frac{3\zeta^{2}}{32} R^{(3)}\approx 0
\end{equation}
get recovered.\\
\hspace*{3mm}Invariance of an action under variation of $\zeta^{\mu} \coloneqq (\zeta, \zeta^{A})$ gives stress tensor (can also be found in \citep{ThirdQuantization})
\begin{eqnarray} \label{StressTensor}
 T^{\nu}_{\mu} \coloneqq \frac{\partial \mathcal{L}}{\partial \left( \frac{\partial \Phi}{\partial \zeta^{\nu}}\right)}\frac{\partial \Phi}{\partial \zeta^{\mu}} - \mathcal{L}\delta_{\mu}^{\nu}
\end{eqnarray}
For now, assume $\zeta$ as time and perform Legendre transformation to get the Hamiltonian
\begin{eqnarray}\label{Hamiltonian}
    & \Pi_\Phi \coloneqq \frac{\partial \left(\sqrt{-G}\mathcal{L}\right)}{\partial \frac{\partial \Phi}{\partial \zeta}} = \sqrt{-G}\frac{\partial \Phi}{\partial \zeta} \\
  &\mathbf{H}_{\Phi} = \int \, \mathscr{D}\zeta^{A}\,\mathcal{H}\\ \nonumber
  &\mathcal{H} \coloneqq \frac{1}{2} \left(\frac{\Pi_\Phi^{2}}{\sqrt{-G}} + \frac{32}{3\zeta^{2}}\sqrt{-G}\partial_A\Phi\bar{G}^{AB}\partial_B\Phi + \sqrt{-G}\mu^2 \Phi^{2}\right)
\end{eqnarray}
The conserved quantity corresponds to $\zeta$ is interpreted as the energy of the field $\Phi$. For $\mu^2<0$, the Hamiltonian would not have a lower bound. Therefore, introduce $\lambda \Phi^{4}$ with some $\lambda>0$ to maintain the Hamiltonian to be positive definite. 
\begin{eqnarray}
  \mathbf{H}_{\Phi} &= \int \, \mathscr{D}\zeta^{A} \, \frac{1}{2}\left( \frac{\Pi_\Phi^{2}}{\sqrt{-G}} + \frac{32}{3\zeta^{2}}\sqrt{-G}\frac{\partial \Phi}{\partial \zeta^{A}}\bar{G}^{AB}\frac{\partial \Phi}{\partial \zeta^{B}}\right)\\ \nonumber 
  &+\int \, \mathscr{D}\zeta^{A} \frac{1}{2}\left( \sqrt{-G}\mu^2 \Phi^{2} + \lambda \Phi^{4} \right)
\end{eqnarray}
The Hamiltonian shows that the quadratic coupling will be non-negative regardless of the signature of the 3-Ricci scalar and justifies the use of $\zeta$ as the time for the geometric scalar field. For $R^{(3)}<0$, $\Phi$ is not necessarily self-coupled. The intrinsic curvature $R^{(3)}>0$ makes the geometric field $\Phi$ self-coupled. \\
\hspace*{5mm}Note: \textit{The Ricci curvature scalar $R^{(3)}$ and cosmological constant $\Lambda$ are not on equal footing. The former one has quadratic coupling, whereas the latter one contributes to the vacuum}.\\
\hspace*{3mm}Although the geometric scalar field satisfies ADM constraints for pure gravity, it is not a pure (or conventional) gravitational field because pure gravitational field have zero stress tensor given by the Einstein field equations $R_{\mu\nu} - \frac{1}{2}g_{\mu\nu}R = 0$. The geometric scalar field has a well-defined, non-trivial stress tensor, and quadratic coupling always remains positive. Every geometric scalar field has a certain $R^{(3)}$. In this sense, it does not live in space-time but has its geometric properties and, the background is dynamic. Intrinsic geometry plays a role similar to the mass in standard special relativistic field theory.
\section{Classical Theory: Applications}
The section shows that the geometric scalar field includes all the properties of gravity. As explained in section 2, the properties of a standard scalar matter get modified. The first subsection shows that $\zeta$ acts as a clock for the classical theory. The second subsection looks at a situation where space-time is static. The last subsection explains gravitational waves and also observes the role of $\mu^2$.
\subsection{FLRW Cosmologies}
\subsubsection{ADM Theory}
The metric tensor for the four dimensional spacetime is given as follows.
\begin{align}
    & g_{\mu\nu} \coloneqq \text{diag} \left(N^2, - q_{ab}\right)\\
    & q_{ab}\coloneqq q(t)\,\text{diag} \left(\frac{1}{1 - k r^2},r^2, r^2 \sin^2\theta\right) 
\end{align}
The intrinsic curvature of the 3-manifold $R^{(3)}=\frac{6k}{q(t)}$. $N(t)$ is the lapse function. The shift vector $N^a$ chosen as 0. The FLRW Hamiltonian is obtained in these coordinates.
\begin{align}\label{ADM_FRW_H}
    H_{ADM} = \frac{q^\frac{1}{2}P_q^2}{3V_0} + 6V_0 k q^\frac{1}{2}
\end{align} 
$V_0 = 4\pi \int dr r^2 \sqrt{1 - k r^2}$ is the volume of fiducial cell and $P_q = 3V_0 \frac{\dot{q}}{N q^\frac{1}{2}}$ is the momentum conjugate to $q$. Now, solving Hamiltonian equations 
\begin{align}
    &\dot{q} = \left\{q, H_{ADM}\right\}_{P.B.} &\dot{P}_q = \left\{P_q, H_{ADM}\right\}_{P.B.}
\end{align}
and using the fact that $H_{ADM} = 0 \rightarrow \frac{\dot{q}^2}{4q} = -k$, we get
\begin{equation}
    \ddot{q} = \frac{\dot{q}^2}{4q} - k = -2k
\end{equation}
For $k= + 1$, the acceleration is negative and $q(t)$ is double-valued function of time `t'. It shows that $q(t)$ cannot be taken as global time in general.
\subsubsection{Geometric Scalar Field}
Now, turn on $P_q \coloneqq -i\frac{\partial}{\partial q}$ and re-interpret the ADM Hamiltonian (\ref{ADM_FRW_H}) as 
\begin{eqnarray}
    \left(\frac{1}{3V_0} \frac{\partial^2}{\partial q^2} - k 6V_0 \right) \Phi = 0
\end{eqnarray}
Following the second section, assume the form of action to be 
\begin{equation}
    \mathscr{A}_{FRW} = \int dq \frac{1}{2} \left(\frac{1}{3V_0}\left(\partial_q \Phi\right)^2 + k 6V_0 \Phi^2\right) 
\end{equation}
The Hamiltonian takes the following form.
\begin{align}
    &\Pi \coloneqq \frac{1}{3V_0} \partial_q \Phi, &\mathbf{H}_{FRW} = \frac{3V_0}{2} \left(\Pi^2 - 2 k \Phi^2\right)
\end{align}
The Hamiltonian equations for the real scalar field are 
\begin{align}
    &\Pi = \left\{\Phi, \mathbf{H}_{FRW}\right\}_{P.B.} &\frac{\partial\Pi}{\partial q} = \left\{\Pi, \mathbf{H}_{ADM}\right\} =  6V_0 \, k\, \Phi
\end{align}
If we interpret $\Phi \sim e^{\pm i q\frac{P_q}{3V_0}}$ as single spacetime-like wave function, this Hamiltonian reduces to the ADM Hamiltonian constraints. For $k=+1$, $q(t)$ is double-valued in the ADM theory. But as discussed in the earlier section, the quadratic coupling is always positive definite for geometric scalar field. Therefore, $q$ plays a role of time in this theory unlike the Wheeler-DeWitt theory.
\subsection{Schwarzschild Geometry}
This subsection shows that the classical geometric scalar field carries singularities existing in the ADM theory.
\subsubsection{ADM Theory}
It is useful to write the Schwarzschild metric in the isotropic radial coordinates for the ADM theory.
\begin{align}
  & g_{\mu\nu} \coloneqq N^2dt^2 - 2N_adx^a dt - q_{ab}dx^a dx^b\\
  & q_{ab} \coloneqq \left(1 + \frac{C}{r}\right)^4 \text{diag} \left( 1,1,1\right) 
\end{align}
The intrinsic curvature $R^{(3)}=0$, the extrinsic curvature tensor $K_{ab} = \frac{1}{2N}\left(q_{ak}D_bN^k + q_{bk}D_aN^k\right)$ and the surface gravity term is
\begin{align}
  \mathscr{A}_{Sch} = -2 \int d^4x \sqrt{q} \left( N\nabla_\mu(K n^\mu) + D_a D^a N\right) 
\end{align}
$\sqrt{q}\coloneqq\sqrt{\text{det }q_{ab}}$, $\nabla_\mu$ is covariant derivative on 4 dimensional manifold such that $\nabla_\mu g_{\nu\rho} = 0$ and $D_a$ is covariant derivative on 3 dimensional manifold such that $D_a q_{bc} = 0$. The Hamiltonian of the surface gravity term in the asymptotic inertial frame (i.e., $N=1$ and $N^a = 0$) gives ADM mass $M_{ADM}$.
\subsubsection{Geometric Scalar Field}
The spatial part of the super-metric is defined in \cite{DeWitt} as
\begin{align}
  &\bar{G}_{AB} = \text{Tr}\left(\mathbf{q}^{-1}\frac{\partial \mathbf{q}}{\partial\zeta^A}\mathbf{q}^{-1}\frac{\partial \mathbf{q}}{\partial\zeta^B}\right) \\ \nonumber
  &\mathbf{q} \coloneqq q_{ab}
\end{align}
 and $\zeta^A \coloneqq f(r) = \left(1+\frac{C}{r}\right)^4$ is used. Because components of the 3-metric are `good coordinates'. The intrinsic curvature $R^{(3)}=0$ implies
\begin{equation}
  \left(\frac{1}{\zeta^D}\frac{\partial}{\partial\zeta}\zeta^D \frac{\partial}{\partial\zeta} - \frac{32}{3\zeta^2}\frac{1}{\sqrt{\bar{G}}}\frac{\partial}{\partial\zeta^A}\sqrt{\bar{G}}\bar{G}^{AB}\frac{\partial}{\partial\zeta^B}\right) \Phi = 0
\end{equation}
But the 3-metric is static and one dimensional. Since $\sqrt{\bar{G}} = \frac{3}{f(r)}$ with $f(r) \coloneqq \left(1 + \frac{C}{r}\right)^4$, the field equation becomes
\begin{equation}
 32 f\frac{\partial}{\partial f}f\frac{\partial}{\partial f} \Phi = 0 \rightarrow \frac{\partial\Phi}{\partial f} = \frac{\alpha}{f}
\end{equation}
$\displaystyle \lim_{f \rightarrow \infty}\Phi = \infty$ is `the black hole singularity'. The conserved quantity corresponds to $f$ is obtained as
\begin{align}
  E = \int df\, T^0_0 = 32 \int \frac{df}{f} f^2 \left(\frac{\partial\Phi}{\partial f}\right)^2 \rightarrow E = \alpha^2 \ln f 
\end{align}
It is interpreted as a potential energy associated with the field $\Phi(f)$.
\subsection{Linearized theory}
I define $q_{ab}\coloneqq\delta_{ab} + h_{ab}$ with $\delta_{ab}>>h_{ab}$. The five dimensional superspace $\zeta^{A} \coloneqq (r_{1},r_{2},\theta_{1},\theta_{2},\theta_{3})$ has three compact and two non-compact coordinates.
\begin{eqnarray}
  \left( \frac{\partial^{2} }{\partial \zeta^{2}} - \frac{\partial^{2}}{\partial \vec{\zeta}^{2}} + \mu^2 \right) \Phi = 0
\end{eqnarray}
The factor $\frac{32}{3}$ is absorbed inside coordinates. The wave equation is asymptotic and it represents a wave passing through surface with intrinsic geometry defined by (\ref{coupling}). Asymptotically flat approximation does not change the intrinsic curvature $R^{(3)}$. It makes field $\Phi$ in a fixed background having intrinsic quantity $\mu^2$. Depending on the intrinsic geometry of the surface, the signal arrives at different times. In this sense, the role of intrinsic curvature is similar to the role of mass in the special relativistic QFTs. But $\mu^2$ is not the mass. Using Iwasawa decomposition $SL(3, \mathbb{R})/SO(3, \mathbb{R})$ can be written in terms of 
\begin{eqnarray}
 &\begin{pmatrix}
  r_{1} &0 &0\\
  0 &r_{2} &0\\
  0 &0 &\frac{1}{r_{1}r_{2}} 
 \end{pmatrix}
 &\begin{pmatrix}
  1 &\theta_{3} &\theta_{2}\\
  0 &1 &\theta_{1}\\
  0 &0 &1 
 \end{pmatrix}
\end{eqnarray}
$r_{1}$ and $r_{2}$ produce deformations along axes by preserving volume. Whereas other three are deformations along respective plane. $\Phi(r_1)$ and $\Phi(r_2)$ represent $+$-polarized waves. Whereas $\Phi(\theta_1)$, $\Phi(\theta_2)$ and $\Phi(\theta_3)$ represent $\times$-polarized waves. 
\section{Quantum Theory}
Quantum geometric effects are assumed to take over the classical geometry near the singularity. Quantization of non-linear classical theory comes with the cost of raising creation and annihilation operators to vectors on coordinate space. An advantage of coordinate space quantization is that, it is possible to relate $(\zeta, \zeta^{A})$ to $(t, \vec{x})$ making interpretation easier.\\
\hspace*{2mm} I define vector-valued annihilation and creation operators
\begin{eqnarray}
  & a_{A} \coloneqq \frac{(-G)^{\frac{1}{4}}}{\sqrt{2}} \biggl( \frac{\Pi}{(-G)^{\frac{1}{4}}} n_{A} + i \sqrt{\frac{32}{3\zeta^{2}}}\frac{\partial \Phi}{\partial \zeta^{A}} + i \omega \Phi n_{A} \biggr)  \\ \nonumber
  & a^{\dagger}_{B} \coloneqq  \frac{(-G)^{\frac{1}{4}}}{\sqrt{2}} \biggl( \frac{\Pi}{(-G)^{\frac{1}{4}}} n_{B} - i\sqrt{\frac{32}{3\zeta^{2}}}\frac{\partial \Phi}{\partial \zeta^{B}} - i \omega \Phi n_{B} \biggr) 
\end{eqnarray}
$n^{A}\coloneqq\frac{\zeta^A}{\sqrt{\bar{G}_{AB}\zeta^A \zeta^B}}$, $(-G)^{\frac{1}{4}} \coloneqq \left( -\text{det } G_{\mu\nu}\right)^\frac{1}{4}$ and $\omega \left( \zeta, \zeta^{A} \right)$ is a solution to the Riccati equation (\ref{Riccati}).
\begin{eqnarray}
  & a^{\dagger}_{A}\bar{G}^{AB}a_{B} + a_{A}\bar{G}^{AB}a_{B}^{\dagger} \\ \nonumber
  & = \frac{\sqrt{-G}}{2}\left( \frac{\Pi^2}{\sqrt{-G}} + \frac{32}{3\zeta^{2}}\frac{\partial \Phi}{\partial \zeta^{A}}\bar{G}^{AB}\frac{\partial \Phi}{\partial \zeta^{B}} + \omega^2 \Phi^2 \right) \\ \nonumber
  & + \omega \sqrt{-G}\sqrt{\frac{32}{3\zeta^{2}}} \, n_{A}\bar{G}^{AC}\left( \frac{\partial \Phi}{\partial \zeta^{C}} \right) \Phi
\end{eqnarray}
The last term can be re-written as 
\begin{eqnarray} \label{SurfaceTerm}
  & \omega \sqrt{-G}\sqrt{\frac{32}{3\zeta^{2}}} \,  n_{A}\bar{G}^{AC} \left( \frac{\partial \Phi}{\partial \zeta^{C}} \right) \Phi \\ \nonumber
  &= \frac{\partial }{\partial \zeta^{C}} \left( \omega \sqrt{-G} \sqrt{\frac{32}{3\zeta^{2}}} \,  n_{A}\bar{G}^{AC}\frac{1}{2} \Phi^{2} \right) \\ \nonumber 
  &- \frac{\partial }{\partial \zeta^{C}} \left( \omega \sqrt{-G} \sqrt{\frac{32}{3\zeta^{2}}} \,  n_{A}\bar{G}^{AC} \right)  \frac{1}{2} \Phi^{2}
\end{eqnarray}
Integral of the first term on the right hand side is surface Integral. Assume it to vanish on the surface. Then
\begin{eqnarray}
  &\omega \sqrt{-G} \sqrt{\frac{32}{3\zeta^{2}}} \,  n_{A}\bar{G}^{AC} \left( \frac{\partial \Phi}{\partial \zeta^{C}} \right) \Phi \\ \nonumber
  & = - \frac{\partial }{\partial \zeta^{C}} \left( \omega \sqrt{-G} \sqrt{\frac{32}{3\zeta^{2}}} \, n^{C} \right)  \frac{1}{2} \Phi^{2}
\end{eqnarray}
Now $\omega \in \mathbb{R}$ is chosen as the solution to following Riccati equation
\begin{eqnarray} \label{Riccati}
 \sqrt{-G} \omega^{2} - \frac{\partial }{\partial \zeta^{C}} \left( \omega \sqrt{-G} \sqrt{\frac{32}{3\zeta^{2}}} \, n_{A}\bar{G}^{AC} \right) = \sqrt{-G}	\mu^2 
\end{eqnarray}
The equation should be solved using `correct' boundary conditions. Such a solution is unique. Examples of such boundary conditions are
\begin{itemize}
  \item FLRW $\kappa=0$ model: $\omega$ that makes the spectrum of the Hamiltonian operator continuous in the limit $q(t) \rightarrow \infty$.
  \item Schwarzschild spacetime: $\omega$ that makes the spectrum of the Hamiltonian operator continuous in the limit $q_{ab}\rightarrow\eta_{ab}$ with $\eta_{ab}$ being flat 3-metric. 
\end{itemize}
Computing the non-trivial commutator,
\begin{eqnarray}
  & a_{A}\bar{G}^{AB}a_{B}^{\dagger} -  a^{\dagger}_{A} \bar{G}^{AB}a_{B} \\ \nonumber
  & = \frac{i}{2} (-G)^{\frac{1}{4}} \sqrt{\frac{32}{3\zeta^{2}}} n_{A}\bar{G}^{AB} \left( \frac{\partial \Phi}{\partial \zeta^{A}}\Pi  - \Pi \frac{\partial \Phi}{\partial \zeta^{A}}\right) \\ \nonumber
  & + \frac{i}{2} \omega (-G)^{\frac{1}{4}} \left(  \Phi \Pi - \Pi \Phi \right)\\ \nonumber
  & = \frac{i}{2} (-G)^{\frac{1}{4}} \sqrt{\frac{32}{3\zeta^{2}}} n_{A}\bar{G}^{AB} \frac{\partial}{\partial \zeta^{A}} \left[ \Phi, \Pi \right] + \frac{i}{2} (-G)^{\frac{1}{4}} \omega  \left[ \Phi, \Pi \right]
\end{eqnarray}
Using property of Dirac delta function $f(x)\delta^\prime (x) = - f^\prime (x)\delta (x)$, we get
\begin{eqnarray}\label{commutator}
  \scriptstyle \left[ a, a^{\dagger} \right] = \frac{\varepsilon_{\text{Pl}}}{2} \left( \sqrt{\frac{32}{3\zeta^{2}}} \partial_C \left((-G)^{\frac{1}{4}}  n^C\right) - (-G)^{\frac{1}{4}}  \omega \right) \delta \left( \vec{\zeta}, {\vec{\zeta^{\prime}}} \right) 
\end{eqnarray}
Since this is the quantization of geometry, the Planck energy $\varepsilon_{\text{Pl}}$ appears in stead of $\hbar$. The above commutator was possible because the inverse metric $\bar{G}^{AB}$ is symmetric. These identities allow us to write the Hamiltonian operator in the discrete space. (i.e. $\int \mathscr{D}\zeta^A\rightarrow\sum_{\zeta^A}$)
\begin{eqnarray} 
  & \mathbf{H}_{\Phi} = \sum_{\zeta^{A}} a^{\dagger}_{A}\bar{G}^{AB}a_{B} \\ \nonumber
  & + \frac{\delta(0)}{2}\varepsilon_{\text{Pl}}\sum_{\zeta^{A}} \left( \sqrt{\frac{32}{3\zeta^{2}}} \partial_C\left((-G)^{\frac{1}{4}} \partial_C n^C\right) - \omega (-G)^{\frac{1}{4}} \right) 
\end{eqnarray}
The second term is the vacuum term. The quantum vacuum is a sea of constantly creating and annihilating geometries. Discarding this term and writing the Hamiltonian operator in terms of number operator $\hat{\textbf{n}}= \sum_A \hat{\textbf{n}}_A$ with ${a^\dagger}^A a_A \coloneqq \left(\sqrt{\frac{32}{3\zeta^{2}}} \partial_C\left((-G)^{\frac{1}{4}}  n^C\right) - \omega (-G)^{\frac{1}{4}}\right) \hat{\textbf{n}}_A$
\begin{eqnarray} \label{HamiltonianOperator}
  \hat{\mathbf{H}}_{\Phi} = \varepsilon_{\text{Pl}} \sum_{\zeta^{A}} \left| \sqrt{\frac{32}{3\zeta^{2}}} \partial_C\left((-G)^{\frac{1}{4}}  n^C\right) - (-G)^{\frac{1}{4}} \omega \right| \hat{\textbf{n}}
\end{eqnarray}
The appearance of the differential equation (\ref{Riccati}) is not surprising. It is a consequence of using coordinate space for quantization. The scalar quantum of the Klein-Gordon field satisfies $\omega^2 = k^2 + m^2$. Similarly, the quantum of the geometric scalar field follows $frequency = \scriptstyle \sqrt{\frac{32}{3\zeta^{2}}} \partial_C\left((-G)^{\frac{1}{4}}  n^C\right) - \omega (-G)^{\frac{1}{4}} $ with $\omega$ being solution to (\ref{Riccati}). $\Phi$ has a single degree of freedom. Therefore the quantum is scalar. $\Pi$ is a collection of creation and destruction operators. But $\Phi$ depends non-linearly on creation and annihilation operators.\\
\hspace*{3mm}\textbf{Momentum:}
The momentum operator defined using stress tensor 
\begin{equation}
  \hat{\mathbf{P}}^{C} = - \frac{32}{3\zeta^2} \sum_{\zeta^{A}}\, \left( \bar{G}^{CB} \frac{\partial \Phi}{\partial \zeta^{B}}\right)\Pi
\end{equation}
does not share eigenstates with the Hamiltonian operator. This is because $\Phi$ depends non-linearly on creation and annihilation operators. Therefore the quantum with a particular 3-metric does not have well-defined momentum at the quantum level.\\
\hspace*{3mm}\textbf{Measurement:}
What is measured is the 3-metric and not the position of the quantum. Assume that the quantum state is `prepared' in the superposition of several 3-geometries. On the measurement, we can observe only one 3-geometry. Quantum mechanically, there is no limit on how fast the quantum can change its 3-geometry. It is because energy eigenstates are not momentum eigenstates. Whereas in the standard quantum field theories in the flat space-time, the quantum with definite energy has definite momentum but does not have a well-defined position.
\subsection{Hilbert Space}
Canonical commutation relations in discrete superspace are given as 
\begin{eqnarray}
  & \left[ a^{\dagger}, a^{\dagger} \right] =  \left[ a, a \right] = 0 \\ \nonumber
  & \left[ a, a^{\dagger} \right] = \beta \, \delta_{\vec{\zeta}, \vec{\zeta^{\prime}}} \hspace*{0.5cm} (\text{say})
\end{eqnarray}
Where $\beta \coloneqq \scriptstyle \frac{\varepsilon_{\text{Pl}}}{2} \left( \sqrt{\frac{32}{3\zeta^{2}}} \partial_C \left((-G)^{\frac{1}{4}}  n^C\right) - (-G)^{\frac{1}{4}}\omega \right)$ is set for simplicity. These commutation relations allow us to construct the Hilbert space following experience of Harmonic oscillator with a major difference. As we move from one geometry (i.e. a particular point in the superspace) to another geometry, $\beta$ changes. The role of creation and annihilation operator get interchanged depending on the signature of $\beta$.\\
\hspace*{3mm}Define vacuum $| 0\rangle$ which gets annihilated by annihilation operator. Assume $\beta >0$,
\begin{eqnarray}
  a_{A} | 0\rangle_A = 0
\end{eqnarray}
any $k$-th state can be obtained by using creation operator
\begin{eqnarray}
  | k_{\zeta, \vec{\zeta}} \rangle^{A} = \frac{1}{\sqrt{k!}} \left({a^\dagger}^{A} (\zeta, \vec{\zeta}) \right)^{k} \, | 0\rangle^A
\end{eqnarray}
The state represents $k$ number of quantum with $(\zeta, \zeta^{A})$. Since creation and annihilation operators are vector-valued in the superspace, $a^\dagger_C \bar{G}^{CA} = {a^\dagger}^A$ is adjoint of $a_A$.  The self-adjointness of the Hamiltonian requires $\omega \in \mathbb{R}$. The state vector can be raised or lowered by the metric $\bar{G}_{AB}$ or the inverse metric $\bar{G}^{AB}$.
\begin{equation}
	| {k_{1}}_{\zeta, \vec{\zeta}}\rangle_A = \bar{G}_{AB} | {k_{1}}_{\zeta, \vec{\zeta}}\rangle^B
\end{equation}
The inner product is defined as 
\begin{eqnarray}
  & _B\langle {k_{2}}_{\zeta, \vec{\zeta^{\prime}}} | {k_{1}}_{\zeta, \vec{\zeta}}\rangle^B = \delta_{k_{1}, k_{2}} \delta \left( \vec{\zeta},\vec{\zeta^{\prime}} \right) 
\end{eqnarray}
Note that $B$ is not summed over in the left hand side of the equation. States $\left\lbrace | k_{\zeta, \vec{\zeta}} \rangle^{A} \right\rbrace$ with $A=1,2,3,4,5$ together describe \textit{single-geometry} and each of the state deals with a particular coordinate in the superspace.
\section{Quantum Theory: Applications}
In this section, I analyze the quantum geometrical aspects of the FLRW models and the Schwarzschild geometry.
\subsection{FLRW Flat Universe}
The Hamiltoian operator (\ref{HamiltonianOperator}) for this case takes the following form
\begin{equation}
	\hat{\mathbf{H}}_{\Phi} = \frac{\varepsilon_{\text{Pl}}}{\sqrt{\zeta}}  \hat{\textbf{n}}
\end{equation}
\hspace*{3mm}The frequency of the quanta varies as $\frac{1}{\sqrt{\zeta}}$. In other words, the frequency gets red-shifted with time. It is the re-statement of expanding the Universe. In the general theory of relativity, a time-like Killing vector field may not always exist. Therefore, defining energy is not straightforward in the general theory of relativity. But in the geometric scalar field, $\zeta$ plays a role of time, and hence corresponding conserved quantity is interpreted as the energy. If we assume that $E_U$ is the energy of the Universe is then, it must have begun at finite non-zero time,
\begin{eqnarray}
  \zeta_{0} = \left(\frac{\varepsilon_{\text{Pl}}}{E_{U}} n\right)^2
\end{eqnarray}
The interpretation in terms of the ADM theory is that the Universe had a non-zero initial volume at the time of the big bang.
\subsection{Schwarzschild Geometry}
Set $\zeta=1$ as the situation is static. The spatial part of the DeWitt metric $\bar{G}_{AB}=\frac{3}{f^2}$ is one dimensional. Because $q_{ab}(\vec{x}) = f(r)\delta_{ij}$. Also the coordinates chosen are isotropic radial coordinates. The unit normal vector $n^A = \frac{f}{\sqrt{3}}$. I choose the trivial solution $\omega = 0$.
\begin{align} 
  \scriptstyle \hat{\mathbf{H}}_{\Phi} &= \scriptstyle \hat{\textbf{n}} \sqrt{\frac{32}{3 \sqrt{3}}} \varepsilon_{\text{Pl}} \sum_{f} \partial_f f^\frac{1}{2} \Rightarrow \hat{\textbf{n}} \sqrt{\frac{32}{3\sqrt{3}}} \varepsilon_{\text{Pl}} \int_{1}^{f} df \partial_f f^\frac{1}{2} \\ \nonumber
  &= \scriptstyle \hat{\textbf{n}} \sqrt{\frac{32}{3\sqrt{3}}}\varepsilon_{\text{Pl}} (f^\frac{1}{2}-1)
\end{align}
The Hamiltonian spectrum is continuous for $f = 1$ justifies the trivial solution. The quantum of energy $E$ can have maximum $\scriptstyle f_{max} = \left( 1 + \sqrt{\frac{3\sqrt{3}}{32}}\frac{E}{\varepsilon_{\text{Pl}}}\right)^2$. To form a black hole $f_{max}\geq f_0$ with $f_0 = (1+1)^4 = 16$ being a value of metric at event horizon. The smallest possible black hole has the energy given as
\begin{eqnarray}
  \scriptstyle E_{min} =  \sqrt{\frac{32}{3\sqrt{3}}}\varepsilon_{\text{Pl}} (f_0^\frac{1}{2}-1) \approx 17 \varepsilon_{\text{Pl}}
\end{eqnarray}
It is the Planck scale black hole. \\
\hspace*{3mm}The stellar black hole of mass $10 M_\odot \approx 10^{39} M_{pl}$ has event horizon at $30 km \approx 10^{39}L_{pl}$. The maximum possible value of the metric $\scriptstyle f_{max} =  \left( 1 + \frac{M}{2 r_{min}}\right)^4 \approx \left( 1 +  10^{40} \right)^2$ and inner radius 
\begin{equation}
	\scriptstyle \frac{10^{39}L_{pl}}{r_{min}} \approx 10^{20} \Rightarrow r_{min} \approx 10^{19}L_{pl} \approx 10^{-16} m
\end{equation} 
in the isotropic radial coordinate system.
\section{Remarks}
\subsection{Issues in quantization of geometry}
\subsubsection{Time}
Kucher (\cite{kucher}, chapter 2) classifies the problem of time in quantum gravity into three. Namely, the problem of functional evolution, the multiple-choice problem, and the Hilbert space problem.
\begin{itemize}
  \item In the ADM theory Hamiltonian constraints obey Poisson bracket $\left\lbrace H(x), H(x^\prime )\right\rbrace \approx 0$. But After raising them to operators on the Hilbert space, we may get $\left[ \hat{H}(x), \hat{H}(x^\prime ) \right] \neq 0$. It is the functional evolution problem. The theory developed in this paper does not have a problem with functional evolution. Because $q_{ab}$ and $P^{ab}$ are not observables in theory. Hamiltonian constraints re-interpreted as a classical field equation. 
  \item Depending on the choice of time, we may get a different quantum theory. In the case of relativistic quantization, the particle's clock serves as the best clock. Similarly, the volume element of the Universe serves as the best clock for the geometric scalar field. Any observer in the Universe can observe the Universe, including the observer inside the black hole. This natural choice of time circumvents the multiple-choice problem.
  \item The third quantized theories face the Hilbert space problem (\cite{kucher}, chapter 11). The quantization of geometric field is not the third quantization. But geometric matter fields are still obtained by re-interpretation. If we follow Kucher's analysis(\citep{kucher}, chapter 11), the problem is deeply rooted in the concept of \textit{particle}. In the third quantized theory, the interpretation of a particle does not change. Since the background is dynamic, it is not clear what is one-particle Hilbert space. But in the geometric scalar field, matter itself is a geometric entity, and \textit{single-geometry} Hilbert space is known. This distinction with the third quantization shows that the geometric scalar field theory does not have a Hilbert space problem. Here, not to get confused with the requirement of a definite solution to the Riccati equation to have a unique Hamiltonian and unique Hilbert space. The latter issue is because of the Riccati equation. Observations give required boundary/initial conditions that solve (\ref{Riccati}).
\end{itemize}
\subsubsection{Semi-classical theory}
While attempting to quantize gravity given by semi-classical Einstein field equations
\begin{equation}
	R_{\mu\nu} - \frac{1}{2}g_{\mu\nu} R  = \frac{8\pi G}{c^4} \left\langle \Psi | \hat{T}_{\mu\nu} | \Psi \right\rangle
\end{equation}
the theory faces several issues. First, the expectation value of the energy-momentum tensor that occurs on the right hand is usually divergent and needs some regularization and re-normalization. During this, counter-terms arise that invoke higher powers of the curvature which may alter field equations fundamentally. Second, these field equations introduce non-linearities that do not go well with the linearity of the quantum theory. Kiefer \citep{KieferBook} has discussed these issues in detail in section 1.2.\\
\hspace*{3mm}The geometric matter field theory views gravity as fundamental and matter fields are geometric entities. In comparison to the field equations above, the right-hand side is 0. The quantum theory is singularity-free. Hence, the theory does not have a regularization issue. A consistent quantum theory exists even though gravity is a non-linear theory. Semi-classical approximation gives $\mathcal{A}_{classical} + \hbar \mathcal{A}_{correction} + \mathscr{O}(\hbar^2)$. On the single geometric re-interpretation $\Psi \sim e^{\pm i P^{ab}q_{ab}}$, the first term gives ADM constraints for pure classical gravity that evolve relative to correction term, effectively making correction part as quantum matter field. 
\subsubsection{Singularities}
For the geometric matter field theory, the space-time itself is the background. The momentum represents the change in geometry. If we quantize the theory in the momentum space, quantum would not have well-defined geometry. Such a situation would have been inconceivable. Therefore unlike standard relativistic quantum field theories, the coordinate space quantization is useful in geometric field theories.\\
\hspace*{3mm}The quantum theory resolves singularities by modifying the metric near a singularity.
\begin{itemize}
	\item The theory predicts the initial finite size of the Universe.
	\item In the case of black hole singularity, the theory sets an upper limit to $f(r)$ as we approach the singularity. But we need further investigation on whether $f(r)$ remains $f_{max}$ for $0\leq r \leq r_{min}$ or it decreases with $r$.
\end{itemize}
It shows that the re-interpretation of the Wheeler-DeWitt equation and implementation of the philosophy of `geometric matter' together solves all issues in the quantization of geometry.
\subsection{Differences and advantages}
The theory has the quantum of corresponding geometric matter. But it does not have the quantum of gravity or graviton.
\begin{enumerate}
	\item \textit{Strings versus Geometric matter}
	\begin{itemize}[leftmargin=*]
		\item Like string theory, the geometric matter field theory also unifies matter fields with gravity. But unlike the string theory, geometric matter field theory does not require extra-dimensions or super-symmetry.
		\item Different vibrating modes of strings give different particles and, gravity arises in the string dynamics. The geometric matter field has geometry being fundamental and, matter fields are geometric entities. 
		\item Unlike string theory, there is no graviton as quanta of gravity in the geometric matter field. 
	\end{itemize}
    \item \textit{Wheeler-DeWitt theory versus geometric matter field theory}
    \begin{itemize}[leftmargin=*]
        \item In the Wheeler-DeWitt theory, $q_{ab}$ and $P^{ab}$ are observables. But the rule $P^{ab}\rightarrow -i \frac{\partial}{\partial q_{ab}}$ does not implement $\text{det }q_{ab}>0$. Therefore, it requires modification. In Geometric scalar field theory, $\Phi(q_{ab})$ and $\Pi_{\Phi}(q_{ab})$ are observables and $q_{ab}$ forms a superspace.
        \item In the Wheeler-DeWitt theory, $q_{ab}$ and $P^{ab}$ are first quantized. In Geometric scalar field theory, $\Phi$ and its conjugate momentum are second quantized.
        \item Wheeler-DeWitt theory quantizes gravity. But it does necessarily unify gravity with matter fields. 
        \item The Wheeler-DeWitt theory for pure gravity does not have global time. One of the coordinates is an internal time. Therefore, not all components are quantized. One of the components of the 3-metric remains classical and, other components get quantized. The situation is mathematically similar to that of the relativistic quantum theory where $t$ among $x^\mu$ remains classical and $\vec{x}$ gets quantized. In Geometric scalar field theory, $(\zeta, \zeta^A)$ form coordinate space, and $\Phi$ gets quantized. $\zeta$ acts as time.
        \item Gravity and matter have a different status in the Wheeler-DeWitt theory. The geometric scalar field makes scalar matter a geometric entity.  
    \end{itemize}
    \item \textit{Third quantized field versus geometric matter}
    \begin{itemize}[leftmargin=*]
        \item Classically, the gravitational field and matter field both defined over four-dimensional spacetime. Quantum mechanically, matter fields are second quantized. Whereas, the gravitational field variables $(\hat{q}_{ab}, \hat{P}^{ab})$ are on the equal footing compared to $(\hat{x}, \hat{p})$. Gravity, unlike the matter fields, comes from the generalization of the special theory of relativity. $$(t, \vec{x}) \xrightarrow{\text{General Relativity}} g_{\mu\nu}(t, \vec{x})$$ \hspace*{3mm} The third quantized fields are defined over  $(q_{ab}, \phi)$. Whereas, geometric matter fields are defined only over $q_{ab}$ or $(\zeta, \zeta^A)$. Geometric matter fields are second quantized. As discussed above, $\Phi(q_{ab}, \phi)$ are dubious because quantum mechanically $q_{ab}$ and $\phi$ are on different levels of quantization.
        \item The domain of the third quantized field is a multiverse. The vacuum of the third quantized field is a sea of creating and annihilating Universes. The geometric scalar field describes scalar matter as a geometric entity. The vacuum of the geometric scalar field is a sea of creating and annihilating virtual quanta of geometry.
        \end{itemize}
         \item \textit{Loop quantum gravity versus geometric matter field theory}
    \begin{itemize}[leftmargin=*]
    	\item The loop quantum gravity is a quantum theory of gravity. It does not necessarily unify matter fields. The geometric matter field is an intrinsically unified theory of geometry and matter fields. 
    	\item The loop variables are first quantized (as discussed above in the third quantized theory part), whereas matter fields are second quantized. The geometric matter field theory does not have such an issue. 
    	\item The loop quantum cosmology has collapsing branch before the big bounce. The geometric matter field theory has a finite size of the Universe at the big bang. But it does not have collapsing branch.
    	\item The loop quantum gravity has the Planck star inside the black hole event horizon. According to the geometric matter field, the metric has an upper limit that depends on the total energy of the black hole. The geometric matter sets a lower limit on the energy of the black hole in the Planck domain.
    \end{itemize}
\end{enumerate}
\hspace*{3mm}There is a standard view that the observation of more and more redshift for farther astronomical objects would mean that the Universe is accelerating. The theory offers a different interpretation than this standard interpretation. The flat FLRW Universe observes more and more redshift for farther astronomical objects is due to the quantum nature of geometry (that is, due to $\frac{1}{\sqrt{\zeta}}$ dependence of frequency.) Note that the model is still a flat FLRW Universe, and it does not have dark energy. Yet frequencies get red-shifted more and more with time. The theory thus explains the observed Universe without the dark energy.
\bibliography{GSF}
\end{document}